\documentclass[a4paper]{amsart}

\usepackage{amsmath,amsthm,bm,mathrsfs}
\usepackage[UKenglish]{babel}
\usepackage{graphicx}
\usepackage{paralist}

\newcommand{\eps}{\varepsilon}
\graphicspath{{./}{figures/}}

\newcommand{\n}{\mathbf{n}}
\newcommand{\R}{\mathbb{R}}

\title{Alzheimer's disease: a mathematical model for onset and progression}
\author[M. Bertsch]{Michiel Bertsch}
\address{Dipartimento di Matematica, Universit\`{a} di Roma ``Tor Vergata'', Via della Ricerca Scientifica 1, 00133 Roma, Italy \newline\indent
Istituto per le Applicazoni del Calcolo ``M. Picone'', Consiglio Nazionale delle Ricerche, Via dei Taurini 19, 00185 Roma, Italy}
\email{bertsch@mat.uniroma2.it}
\author[B. Franchi]{Bruno Franchi}
\address{University of Bologna, Department of Mathematics, Piazza di Porta S. Donato 5, 40126 Bologna, Italy}
\email{bruno.franchi@unibo.it}
\author[N. Marcello]{Norina Marcello}
\address{S.C. Neurologia, Azienda Ospedaliera Santa Maria Nuova-IRCCS, Viale Risorgimento 80, 42123 Reggio Emilia, Italy}
\email{norina.marcello@asmn.re.it}
\author[M. C. Tesi]{Maria Carla Tesi}
\address{University of Bologna, Department of Mathematics, Piazza di Porta S. Donato 5, 40126 Bologna, Italy}
\email{mariacarla.tesi@unibo.it}
\author[A. Tosin]{Andrea Tosin}
\address{Istituto per le Applicazioni del Calcolo ``M. Picone'', Consiglio Nazionale delle Ricerche, Via dei Taurini 19, 00185 Roma, Italy}
\email{andrea.tosin@polito.it}

\begin{document}



\begin{abstract}
In this paper we propose a mathematical model for the onset and progression of Alzheimer's disease based on transport and diffusion equations. We regard brain neurons as a continuous medium, and structure them by their degree of malfunctioning. Two different mechanisms are assumed to be relevant for the temporal evolution of the disease:
\begin{inparaenum}[i)]
\item diffusion and agglomeration of soluble polymers of amyloid, produced by damaged neurons;
\item neuron-to-neuron prion-like transmission.
\end{inparaenum}
We model these two processes by a system of Smoluchowski equations for the amyloid concentration, coupled to a kinetic-type transport equation for the distribution function of the degree of malfunctioning of neurons. The second equation contains an integral term describing the random onset of the disease as a jump process localised in particularly sensitive areas of the brain. 
Our numerical simulations are in good qualitative agreement with clinical images of the disease distribution in the brain which vary from early to advanced stages.
\end{abstract}

\maketitle

\section{Introduction}
Alzheimer's disease (AD) is one of the most common late life dementia, with huge social and economic impact~\cite{Blennow_et_al,hurd_et_al,mattson}. Its global prevalence, about $24$ millions in $2011$, is expected to double in $20$ years~\cite{reitz_etal_epidemiology}. \emph{In silico} research, based on mathematical modelling and computer simulations~\cite{AFMT,Cruz08071997,EDELSTEINKESHET2002301,good_murphy,helal_et_al,Murphy_Pallitto,raj,Urbanc19991330}, effectively supplements \emph{in vivo} and \emph{in vitro} research. We present a multiscale model for the onset and evolution of AD which accounts for the diffusion and agglomeration of amyloid-$\beta$ (A$\beta$) peptide (amyloid cascade hypothesis~\cite{haass_delkoe,karran_et_al}), and the spreading of the disease through neuron-to-neuron transmission (prionoid hypothesis~\cite{Braak_DelTredici}). 

Indeed, to cover such diverse facets of AD in a single model, different spatial and temporal scales must be taken into account: microscopic spatial scales to describe the role of the neurons, macroscopic spatial and short temporal (minutes, hours) scales for the description of relevant diffusion processes in the brain, and large temporal scales (years, decades) for the description of the global development of AD. The way in which we combine distinct scales in a single model forms the core and major novelty of the paper.

Following closely the biomedical literature on AD, we briefly describe the processes which we shall include in our model. In the neurons and their interconnections several \emph{microscopic phenomena} take place. It is largely accepted that beta amyloid (A$\beta$), especially its  highly toxic oligomeric isoforms A$\beta_{40}$ and A$\beta_{42}$, play an important role in the process of the cerebral damage (the so-called \emph{amyloid cascade hypothesis}~\cite{karran_et_al}). In this note we focus on  the role of A$\beta_{42}$ in its \emph{soluble} form, which recently has been suggested to be the principal cause of neuronal death and eventually dementia~\cite{walsh}. At the level of the neuronal membrane, monomeric A$\beta$ peptides originate from the proteolytic cleavage of a transmembrane glycoprotein, the amyloid precursor protein (APP). By unknown and partially genetic reasons, some neurons present an unbalance between produced and cleared A$\beta$ (we refer to such neurons as damaged neurons). In addition to this, it has been proposed that neuronal damage spreads in the neuronal net through a neuron-to-neuron prion-like propagation mechanism~\cite{Braak_DelTredici,raj}.

On the other hand, \emph{macroscopic phenomena} take place at the level of the cerebral tissue. The monomeric A$\beta$ produced by damaged neurons diffuses through the microscopic tortuosity of the brain tissue and undergoes a process of agglomeration, leading eventually to the formation of  long, insoluble amyloid fibrils, which accumulate in spherical deposits known as senile plaques. Moreover, soluble A$\beta$ shows a multiple neurotoxic effect: it induces a general inflammation that activates the microglia (the resident immune cells in the central nervous system) which in turn secretes proinflammatory innate cytokines~\cite{griffith_etal_cytokine} and, at the same time, increases intracellular calcium levels~\cite{good_murphy} yielding ultimately apoptosis and neuronal death.

The model we present is a conceptual interdisciplinary construction based on clinical and experimental evidence, yielding in particular numerical simulations and related graphs, that can be compared with time-dependent trajectories of AD biomarkers (see e.g.,~\cite{jack_et_al_1,jack_et_al_2}). In particular, Figure~\ref{fig:beta=1} fits the core of the model proposed in~\cite{jack_et_al_1} for the temporal progression of the abnormalities in AD biomarkers, which identifies two subsequent periods: 
\begin{itemize} 
\item[$-$] a first period of $\beta$-amyloidosis characterised prevalently by reductions
in CSF $A\beta_{42}$ and increased amyloid plaque formation (biomarkers of this first period in our model correspond to CSF-A$\beta_{42}$ and PIB-PET, Pittsburgh compound B - Positron Emission Tomography); 
\item[$-$] a second one characterised by neuronal dysfunction and neurodegeneration (for this period we only take into account the structural MRI). 
\end{itemize}

Of particular medical interest is the initial stage of the second period, which is commonly referred to as  Mild Cognitive Impairment (MCI): see e.g.,~\cite{petersen_et_al}.

\section{Mathematical model}
Highly toxic oligomeric isoforms of beta amyloid, A$\beta_{40}$ and A$\beta_{42}$, cause cerebral damage. Here we restrict our attention to A$\beta_{42}$ (shortly A$\beta$ in the sequel) in \emph{soluble} form, generally considered the principal cause of neuronal death and dementia~\cite{walsh}. Monomeric A$\beta$ peptides originate from proteolytic cleavage of a transmembrane glycoprotein, the amyloid precursor protein (APP). In AD, neurons progressively present an unbalance between produced and cleared A$\beta$, but the underlying mechanism is still largely unknown. On the other hand it was proposed that neuronal damage spreads in the neural pathway through a neuron-to-neuron prion-like propagation mechanism~\cite{Braak_DelTredici,raj}.

Soluble A$\beta$ diffuses through the microscopic tortuosity of the brain tissue and undergoes an agglomeration process. Eventually this leads to the formation of long, insoluble fibrils, accumulating in spherical deposits known as senile plaques. Soluble A$\beta$ has a multiple neurotoxic effect~\cite{good_murphy,griffith_etal_cytokine}. In our model we do not enter the details of the brain tissue, we neglect the action of the $\tau$-protein, we simplify the role of microglia and neglect its multifaceted mechanism (see e.g.,~\cite{EDELSTEINKESHET2002301} and~\cite{Straughan05}). We simply assume that high levels of soluble amyloid are toxic for neurons.

We identify a portion of the cerebral tissue with a 3-dimensional region $\Omega$ and $x\in\Omega$ indicates a generic point. Two temporal scales are needed to simulate the longitudinal evolution of the disease over a period of years: a short (i.e., rapid) $s$-scale (unit time coincides with hours) for the diffusion and agglomeration of A$\beta$~\cite{Meyer-Luhmann_nature}, and a long (i.e., slow) $t$-scale (unit time coincides with several months) for the progression of AD, so $\Delta{t}=\eps\Delta s$ for a small constant $\eps\ll 1$.

We denote the molar concentration of soluble A$\beta$ polymers of length $m$ at point $x$ and time $s$ by $u_m(x,\, s)$, with $1\leq m<N$. That of clusters of oligomers of length $\geq N$ (fibrils) is denoted by $u_N(x,\,s)$ and may be thought as a medical parameter (the plaques), clinically observable through PIB-PET (~\cite{Nordberg}).

To model the aggregation of A$\beta$ $m$-polymers ($1<m<N$) we follow~\cite{AFMT},
\begin{center}
	[variation in (short) time] $=$ [diffusion] $+$ [agglomeration],
\end{center}
which, in mathematical terms, leads to the Smoluchowski equation with diffusion:
\begin{equation}\label{amyloid equation}
	\partial_s u_m=d_m\nabla^2u_m +\left[\dfrac{1}{2}\sum_{j=1}^{m-1}a_{j,m-j}u_ju_{m-j}
		-u_m\sum_{j=1}^{N} a_{m,j} u_j\right].
\end{equation}
where $d_m>0$, $m=1,\,\dots,\,N$, and $a_{i,j}=a_{j,i}>0$, $i,\,j=1,\,\dots,\,N$ (giving the factor $\frac12$ in \eqref{amyloid equation}).

We refer to~\cite{AFMT,FT_torino} for an extensive discussion of~\eqref{amyloid equation}. For reasons related to the model, we can assume that the diffusion coefficients $d_m$ are small when $m$ is large, since big assemblies do not move. In fact, the diffusion coefficient of a soluble peptide scales approximately as a reciprocal of the cube root of its molecular weight (see~\cite{Goodhill1997} and also~\cite{Nicholson1998207}).

Applications of the Smoluchowski equation to the description of the agglomeration of A$\beta$ amyloid appear in~\cite{Murphy_Pallitto}. In this paper, the authors compare experimental data, obtained \emph{in vitro}, with numerical simulations based on the Smoluchowski equation (without diffusion) in order to describe the process leading to insoluble fibril aggregates from soluble amyloid. The form of the coefficients $a_{i,j}$ (the coagulation rates) we use has been considered by \emph{in vitro} Murphy \& Pallitto (see~\cite{Murphy_Pallitto} and~\cite{Pallitto_Murphy_2001}). According to formula (13) in~\cite{Murphy_Pallitto}, the coagulation rates \emph{in silico} in our equations take the form
\begin{equation}\label{formula 13}
a_{i,j}  = \mathrm{const.}\, 
\dfrac{1}{i+j} \cdot
\left( \dfrac{\ln(i/d)+\nu_i}{i}+ \dfrac{\ln(j/d)+\nu_j}{j} \right),
\end{equation}
where $i,j$ are the lengths of the fibrils, $d$ is their diameter, and $\nu_i = 0.312 + 0.565(i/d)^{-1} - 0.1(i/d)^{-2}$. 

The physical arguments leading to formula \eqref{formula 13} rely on sophisticated statistical mechanics considerations (see also~\cite{tomsky_murphy}).

Since $d$ can be assumed very small, without loss of generality we can assume $\nu_i=\nu$ for $=1,\dots, N-1$. Thus we can replace \eqref{formula 13} by
\begin{equation}\label{formula 13 bis}
a_{i,j}  = \mathrm{const.}\, 
\dfrac{1}{i+j} \cdot
\left( \dfrac{\ln(i/d)+\nu}{i}+ \dfrac{\ln(j/d)+\nu}{j} \right) = \dfrac{1}{ij} \cdot \big(\nu + |\ln d| + O(\ln N)\big).
\end{equation}
Since $N$ is finite, in our numerical simulations we use a slightly approximate form of these coefficients, taking
\begin{equation}\label{aij}
	a_{i,j}=\alpha\frac{1}{ij}, \quad \text{where\ } \alpha>0.
\end{equation}

Smoluchowski equations with diffusion have already been considered in the literature (without reference to A$\beta$ amyloid and Alzheimer's disease) with diverse boundary conditions: see for instance~\cite{drake} for a general introduction, and~\cite{Laurencot_Mischler_revista,amann,wrzosek,amann2,amann3}.

Neurons produce A$\beta$ monomers, whence the equation for $u_1$ contains a \emph{source term} $\mathcal{F}$:
\begin{equation}\label{amyloid equation: m=1}
	\partial_s u_1=d_1\nabla^2u_1-u_1\sum_{j=1}^{N}a_{1,j}u_j+\mathcal{F}.
\end{equation}
Since fibrils are assumed  not to move, the equation for $u_N$ has no diffusion term, and takes the form (see (4) in~\cite{AFMT}):
\begin{equation}\label{eq m=N} 
	\partial_s u_N=\frac{1}{2}\sum_{\substack{j+k\geq N \\ k,\,j<N}}a_{j,k}u_ju_{k}.
\end{equation}
It is coherent with experimental data to assume $a_{N,N}=0$ for large $N$. This is equivalent to saying that large oligomers do not aggregate with each other.

The justification of the condition $j,k<N$ in \eqref{eq m=N} requires a few more words. In fact, we must remember that the meaning of $u_N$ differs from that of $u_m$, $m<N$, as well as the identity
\begin{equation}\label{spiegazione 1}
\frac12 \sum_{ j+k\ge N, k<N,j<N} a_{j,k}u_ju_{k} =
\frac12 \sum_{ j+k\ge N} a_{j,k}u_ju_{k} -u_N\sum_{j=1}^Na_{N,j}u_j.
\end{equation}
The idea is that $u_N$ should describe the sum of the densities of all the ``large'' assemblies. We assume that large assemblies exhibit all the same coagulation properties and do not coagulate with each other. Let us briefly show how \eqref{eq m=N} is obtained: we start by writing the exact Smoluchowski equation for all $ m\ge 1 $  using $\tilde u_m$ instead of $u_m$ in order to avoid confusion, i.e. nothing but the PDE in \eqref{amyloid equation} with $m$ ranging from $2$ to $\infty$. We have
\begin{equation}\label{per yves 1}
\frac{\partial }{\partial t}\tilde u_m = d_m\nabla^2\tilde u_m 
- \tilde u_m\sum_{j=1}^{N} a_{m,j} \tilde u_j
 + \frac12 \sum_{j=1}^{m-1} a_{j,m-j}\tilde u_j \tilde u_{m-j},
\end{equation}
where, coherently with our assumptions, we assume
\begin{itemize}
\item[i)] $d_m=d_N$ for $m\ge N$;
\item[ii)] $a_{m,j}=a_{N,j}$ for $m\ge N$. In particular, if $m,j\ge N$, $a_{m,j}=a_{N,j}=a_{N,N}=0$.
\end{itemize}
Therefore, if $m\ge N$, (\ref{per yves 1}) becomes 
\begin{equation}\label{per yves 2}
\frac{\partial }{\partial t}\tilde u_m = d_N\nabla^2\tilde u_m 
- \tilde u_m\sum_{j=1}^{N-1} a_{N,j} \tilde u_j
 + \frac12 \sum_{j=1}^{m-1} a_{j,m-j}\tilde u_j \tilde u_{m-j},
\end{equation}
Now we sum up \eqref{per yves 2} for $m\ge N$, and we set for a while $v:=\sum_{m\ge N} \tilde u_m$. We want to show precisely that $v$ satisfies the equation \eqref{eq m=N} (satisfied by  $u_N$). By i), we have
\begin{equation*}
\begin{split}
\dfrac{\partial v}{\partial t} 
& = d_N\nabla^2 v
- \sum_{m\ge N}\tilde u_m \sum_{j=1}^{N-1} a_{N,j} \tilde u_j
+\frac12  \sum_{m\ge N}\sum_{i=1}^{m-1} a_{i,m-i}\tilde u_i \tilde u_{m-i} 
\\&:= 
 d_N\nabla^2 v -I_1 + \frac12  I_2.
\end{split}
\end{equation*}
It is clear that 
$$ I_1= \sum_{m\ge N}\tilde u_m \sum_{j=1}^{N-1} a_{N,j}\tilde u_j = v \sum_{j=1}^{N-1} a_{N,j}\tilde u_j, $$
that is precisely the second term in \eqref{spiegazione 1}, since $a_{N,N}=0$. As for $I_2$, if we set $j:=i$ and $k:=m-i$, we obtain the first term in  \eqref{spiegazione 1}. Finally, if set  $u_m=\tilde u_m$ for $m<N$ and $u_N=v$ we recover the PDE in (\ref{eq m=N}), as desired.

We model the \emph{degree of malfunctioning} of a neuron with a parameter $a$ ranging from $0$ to $1$: $a$ close to $0$ stands for ``the neuron is healthy'' whereas $a$ close to $1$ for ``the neuron is dead''. This parameter, although introduced for the sake of mathematical modelling (see also~\cite{raj}), can be compared with medical images from Fluorodeoxyglucose PET (FDG-PET~\cite{mosconi_et_al}).

Given $x\in\Omega$, $t\geq 0$, and $a\in [0,\,1]$,
\begin{equation*}
	f(x,\,a,\,t)\,da
\end{equation*}
indicates the \emph{fraction} of neurons close to $x$ with degree of malfunctioning at time $t$ between $a$ and $a+da$. The progression of AD occurs at the long time scale $t$, over decades, and is determined by the \emph{deterioration rate}, $v=v(x,\,a,\,t)$, of the health state of the neurons:
\begin{equation}\label{health macro}
	\partial_t f+\partial_a(fv[f])=0.
\end{equation}
Here $v[f]$ indicates that the deterioration rate depends on $f$ itself. The onset of AD will be included in a subsequent step.

We assume that 
\begin{equation}\label{velocity}
	v[f]=\iint_{\Omega\times[0,\,1]}\mathcal{K}(x,\,a,\,y,\,b)f(y,\,b,\,t)\,dy\,db
		+\mathcal{S}(x,\,a,\,u_1(x,\,s),\,\dots,\,u_{N-1}(x,\,s)).
\end{equation}
The integral term describes the possible prion-like propagation of AD through the neural pathway. Malfunctioning neighbours are harmful for a neuron's health state, while healthy ones are not:
\begin{align*}
	\mathcal{K}(x,\,a,\,y,\,b) &\geq 0 &\forall\,x,\,y\in\Omega,\ a,\,b\in[0,\,1], \\
	\mathcal{K}(x,\,a,\,y,\,b) &= 0 & \text{if\ } a>b.
\end{align*}
Typically  
\begin{equation*}
	\mathcal{K}(x,\,a,\,y,\,b)=\mathcal{G}(x,\,a,\,b)H(x,\,y)
\end{equation*}
with, for example,
\begin{equation*}
	\mathcal{G}(x,\,a,\,b)=C_{\mathcal G}(b-a)^{+}, \qquad H(x,\,y)=h(\vert x-y\vert),
\end{equation*}
where $(\cdot)^{+}$ denotes the positive part ($x^+:=\max\{0,\,x\}$) while $h(r)$ is a nonnegative and decreasing function, which vanishes at some $r=r_0$ and satisfies $\int_{\vert y\vert<r_0}h(\vert y\vert)\,dy=1$. In the limit $r_0\to 0$,~\eqref{velocity} reduces to
\begin{equation}\label{velocity bis}
	v[f]=\int_0^1\mathcal{G}(x,\,a,\,b)f(x,\,b,\,t)\,db+\mathcal{S}(x,\,a,\,u_1(x,\,s),\,\dots,\,u_{N-1}(x,\,s)).
\end{equation}
Since we aim at a minimal effective model, we avoid precise assumptions on the underlying biological processes expressed by $\mathcal{K}$.

The term $\mathcal{S}\geq 0$ in~\eqref{velocity} and~\eqref{velocity bis} models the action of toxic A$\beta$ oligomers, ultimately leading to apoptosis. For example
\begin{equation}\label{mathcal S}
	\mathcal{S}=C_{\mathcal S}(1-a){\left(\sum\limits_{m=1}^{N-1}mu_m(x,\,s)-\overline{U}\right)}^{+}
\end{equation}
The threshold $\overline{U}>0$ indicates the minimal amount of toxic A$\beta$ needed to damage neurons, assuming that the toxicity of soluble $A\beta$-polymers does not depend on $m$. In reality length dependence has been observed~\cite{Ono_et_al}, but, to our best knowledge, quantitive data are only available for very short molecules (see~\cite[Table 2]{Ono_et_al}). For long molecules any analytic expression would be arbitrary.

Since A$\beta$ monomers are produced by neurons and the production increases if neurons are damaged, we choose in~\eqref{amyloid equation: m=1}
\begin{equation}\label{mathcal F}
	\mathcal{F}=\mathcal{F}[f]=C_{\mathcal F}\int_0^1(\mu_0+a)(1-a)f(x,\,a,\,t)\,da.
\end{equation}
The small constant $\mu_0>0$ accounts for A$\beta$ production by healthy neurons (dead neurons do not produce amyloid). 

To describe the onset of AD we assume that in small, randomly chosen parts of the cerebral tissue, concentrated for instance in the hippocampus, the degree of malfunctioning of neurons randomly jumps to higher values due to external agents or genetic factors. This leads to an additional term in the equation for $f$, 
\begin{equation*}
	\partial_t f+\partial_a\left(fv[f]\right)=J[f],
\end{equation*}
where
\begin{equation}\label{J formula}
	J[f]=\eta\left(\int_0^1 P(t,\,a_\ast\to a)f(x,\,a_\ast,\,t)\,da_\ast-f(x,\,a,\,t)\right)\chi(x,\,t).
\end{equation}
$P(t,\,a_\ast\to a)$ is the probability to jump from state $a_\ast$ to state $a\in[0,\,1]$ (obviously, $P(t,\,a_\ast\to a)=0$ if $a<a_\ast$), $\chi(x,\,t)$ describes the random jump distribution, and $\eta$ is the jump frequency. In most of our numerical tests we choose 
\begin{equation*}
	P(t,\,a_\ast\to a)\equiv P(a_\ast\to a)=
		\begin{cases}
			\dfrac{2}{1-a_\ast} & \text{if\ } a_\ast\leq a\leq\frac{1+a_\ast}{2} \\
			0 & \text{otherwise},
		\end{cases}
\end{equation*}
i.e., we neglect randomness and we set $\chi(x,\,t)\equiv\chi(x)$ concentrated in the hippocampus. For a simulation with a random jump distribution, see Figure~\ref{fig:disease_sorgenti_casuali}.

To model the phagocytic activity of the microglia as well as other bulk clearance processes~\cite{Iliff_et_al}, we add to~\eqref{amyloid equation} and~\eqref{amyloid equation: m=1} a term $-\sigma_mu_m$, where $\sigma_m>0$. This leads to the system
\begin{equation}\label{complete system}
	\begin{cases}
		\partial_t f+\partial_a\left(fv[f]\right)=J[f] \\
		\eps\partial_tu_1=d_1\nabla^2u_1-u_1\sum\limits_{j=1}^{N}a_{1,j}u_j+\mathcal{F}[f]-\sigma_1u_1 \\
		\eps\partial_tu_m=d_m\nabla^2u_m+\dfrac{1}{2}\sum\limits_{j=1}^{m-1}a_{j,m-j}u_ju_{m-j} \\
		\phantom{\eps\partial_tu_m=d_m\nabla^2u_m}-u_m\sum\limits_{j=1}^{N}a_{m,j}u_j-\sigma_m u_m & (2\leq m<N) \\
		\eps\partial_tu_N=\dfrac{1}{2}\sum\limits_{\substack{j+k\geq N \\ k,\,j<N}}a_{j,k}u_ju_{k},
	\end{cases}
\end{equation}
where $v[f]$ is given by~\eqref{velocity} or~\eqref{velocity bis} (with $s$ replaced by $\eps^{-1}t$), $\mathcal{F}[f]$ by~\eqref{mathcal F}, and $J[f]$ by~\eqref{J formula}. Since we are interested in longitudinal modelling, we assume that initially, at $t=0$, the brain is healthy, with a small uniform distribution of soluble amyloid.

\section{Problem setting and discretisation of the equations}
In this section we detail the structure of the domain and the boundary conditions which we will use, in the next Section~\ref{sec:num_res}, to produce numerical simulations. We also discuss the discretisation of the equations~\eqref{complete system}.

\subsection{Physical domain and boundary conditions}
\begin{figure}[t]
\centering
\includegraphics[scale=0.25]{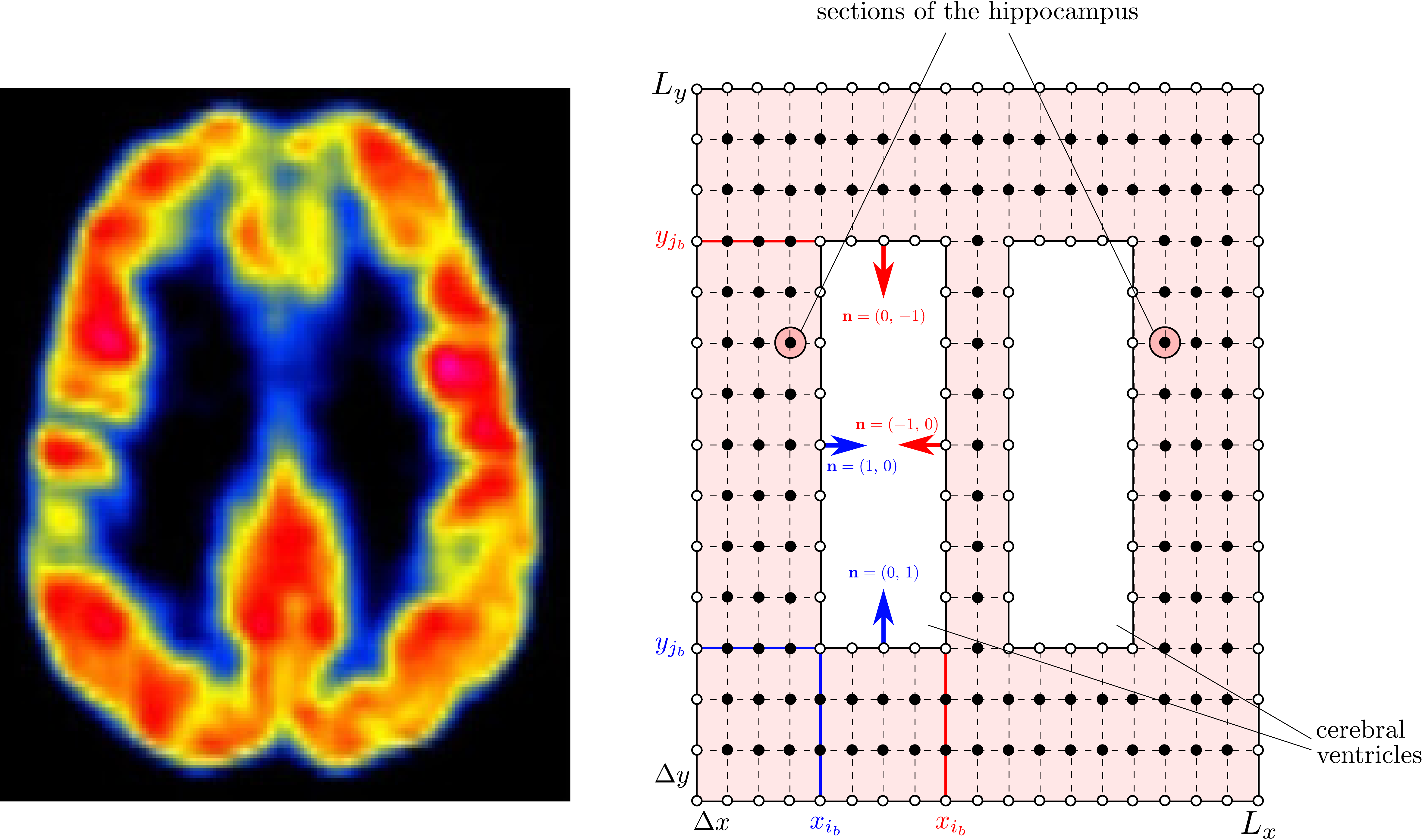}
\caption{Left: A real transverse section of the brain (reproduced from~\cite{miller2006RR} with kind permission of the publisher). Right: Two-dimensional schematisation for numerical purposes. Black dots are the internal nodes of the numerical grid, where discretised equations are solved, while white dots are boundary nodes, where boundary conditions are imposed.}
\label{fig:domain}
\end{figure}

We consider the two-dimen\-sional transverse section of the brain illustrated in Fig.~\ref{fig:domain}. Since approximating a real brain section is a quite complicated issue, for the sake of simplicity we schematise it as a box $\Omega\subset\R^2$, $\Omega=[0,\,L_x]\times [0,\,L_y]$, with two inner rectangular holes representing the sections of the cerebral ventricles. We also identify, close to the front part of the ventricles, the two sections of the hippocampus, which we represent as two small circles. Unlike the cerebral ventricles, the sections of the hippocampus are meant as actual portions of the domain $\Omega$, not as holes.

On the outer boundary of $\Omega$, say $\partial\Omega_\text{out}$, we assume vanishing normal polymer flow. Therefore we impose a homogeneous Neumann condition for the diffusing amyloid polymers:
\begin{equation}
	-\frac{d_m}{\epsilon}\nabla{u_m}\cdot\n=0 \quad \text{on } \partial\Omega_\text{out}, \quad
		m=1,\,\dots,\,N-1,
	\label{eq:bc_Neumann}
\end{equation}
$\n$ being the outward normal unit vector to $\partial\Omega_\text{out}$. Notice that no boundary condition is required for the concentration $u_N$ of the fibrillar amyloid, since its equation does not feature space dynamics (cf. the last equation in~\eqref{complete system}).

On the inner boundary of $\Omega$, say $\partial\Omega_\text{in}$, that is the boundaries of the cerebral ventricles, we model the removal of A$\beta$ from cerebrospinal fluid (CSF) through the choroid plexus (cf.~\cite{Iliff_et_al,serot_et_al}) by an outward polymer flow proportional to the concentration of the amyloid. For this, we impose a Robin boundary condition of the form:
\begin{equation}
	-\frac{d_m}{\epsilon}\nabla{u_m}\cdot\n=\beta u_m \quad \text{on } \partial\Omega_\text{in}, \quad
		m=1,\,\dots,\,N-1,
	\label{eq:bc_Robin}
\end{equation}
with $\beta>0$ a constant.

We discretise $\Omega$ by means of a two-dimensional structured orthogonal grid, whose points have coordinates $x_{i,j}=(x_i,\,y_j)=(i\Delta{x},\,j\Delta{y})$ with $\Delta{x}=L_x/N_x$, $\Delta{y}=L_y/N_y$, $N_x$, $N_y$ being the numbers of discretisation points in the $x$ and $y$-direction, respectively, and $i=0,\,\dots,\,N_x$, $j=0,\,\dots,\,N_y$. See Fig.~\ref{fig:domain}. We also introduce a time lattice $t_n=n\Delta{t}$, $n=0,\,1,\,2,\,\dots$.

Letting $(u_m)_{i,j}^n\approx u_m(x_{i,j},\,t_n)$ denote an approximation of the concentration of the $m$-polymers of amyloid in the point $x_{i,j}\in\Omega$ at time $t_n$, on the numerical grid the Neumann boundary condition~\eqref{eq:bc_Neumann} becomes simply:
\begin{gather*}
	\left.
	\begin{array}{r}
		\left.
		\begin{array}{r}
			(u_m)^n_{0,j}=(u_m)^n_{1,j} \\
			(u_m)^n_{N_x,j}=(u_m)^n_{N_x-1,j}
		\end{array}
		\right\} \quad j=1,\,\dots,\,N_y-1 \\[5mm]
		\left.
		\begin{array}{r}
			(u_m)^n_{i,0}=(u_m)^n_{i,1} \\
			(u_m)^n_{i,N_y}=(u_m)^n_{i,N_y-1}
		\end{array}
		\right\} \quad i=1,\,\dots,\,N_x-1
	\end{array}
	\right\} \quad m=1,\,\dots,\,N-1.
\end{gather*}

Concerning the Robin boundary condition~\eqref{eq:bc_Robin}, we discretise the components of the gradient via the forward Euler formula, then we take into account the orientation of the vector $\n$ as indicated in Fig.~\ref{fig:domain} to find:
\begin{itemize}
\item[$-$] along the left boundary of each cerebral ventricle
$$ (u_m)^n_{i_b,j}=\frac{(u_m)^n_{i_b-1,j}}{1+\epsilon\beta\Delta{x}/d_m}, $$
where $i_b\in\{0,\,\dots,\,N_x\}$ denotes the grid index in the $x$-direction such that $x_{i_b}=i_b\Delta{x}$ is the abscissa of the boundary;
\item[$-$] along the right boundary of each cerebral ventricle
$$ (u_m)^n_{i_b,j}=\frac{(u_m)^n_{i_b+1,j}}{1+\epsilon\beta\Delta{x}/d_m}; $$
\item[$-$] along the lower boundary of each cerebral ventricle
$$ (u_m)^n_{i,j_b}=\frac{(u_m)^n_{i,j_b-1}}{1+\epsilon\beta\Delta{y}/d_m}, $$
where $j_b\in\{0,\,\dots,\,N_y\}$ denotes the grid index in the $y$-direction such that $y_{j_b}=j_b\Delta{y}$ is the ordinate of the boundary;
\item[$-$] along the upper boundary of each cerebral ventricle
$$ (u_m)^n_{i,j_b}=\frac{(u_m)^n_{i,j_b+1}}{1+\epsilon\beta\Delta{y}/d_m}. $$
\end{itemize}

\subsection{Discretisation of the Smoluchowski equations}
In order to approximate the equations for the $u_m$'s, $m=1,\,\dots,\,N-1$, we use a fractional step procedure in time: first we solve the diffusion and reaction parts, then we add the coagulation and possibly the source (for $u_1$) parts.

Adopting a Finite Difference discretisation of the Laplace operator $\nabla^2$ we obtain the scheme:
\begin{equation*}
	\begin{cases}
		(u_m)^\ast_{i,j}=(u_m)^n_{i,j}+
			\frac{\Delta{t}}{\epsilon}\left(d_m\frac{(u_m)^n_{i-1,j}-
				2(u_m)^n_{i,j}+(u_m)^{n}_{i+1,j}}{\Delta{x}^2}\right. \\[3mm]
				\hspace{4cm}\left.+d_m\frac{(u_m)^n_{i,j-1}-2(u_m)^n_{i,j}+(u_m)^n_{i,j+1}}{\Delta{y}^2}-\sigma_m(u_m)^n_{i,j}\right) \\[3mm]
		(u_m)^{n+1}_{i,j}=(u_m)^\ast_{i,j}+\frac{\Delta{t}}{\epsilon}\left(\frac{1}{2}
			\sum\limits_{h=1}^{m-1}a_{h,m-h}(u_h)^\ast_{i,j}(u_{m-h})^\ast_{i,j}\right. \\[3mm]
				\hspace{4cm}\left.-(u_m)^\ast_{i,j}\sum\limits_{h=1}^{N}a_{m,h}(u_h)^\ast_{i,j}\right),
	\end{cases}
\end{equation*}
where $(u_m)^\ast_{i,j}$ denotes the temporary solution computed after the first fractional time step. For an alternative Finite Element discretisation of Smoluchowski equations see e.g.,~\cite{AFMT}.

The scheme above applies to all inner nodes $x_{i,j}$ of the numerical grid (that means $1\leq i\leq N_x-1$, $1\leq j\leq N_y-1$ excluding furthermore the indexes $i_b,\,j_b$ identifying the inner boundary $\partial\Omega_\text{in}$) and to the A$\beta$ $m$-polymers with $m=2,\,\dots,\,N-1$. Because of the adopted approximation of the diffusion part, the following constraint on the time and space steps has to be enforced for the stability of the numerical scheme:
\begin{equation}
	\Delta{t}\leq\frac{\epsilon}{4}\cdot\frac{\min\{\Delta{x}^2,\,\Delta{y}^2\}}{\max\limits_{1\leq m\leq N-1}d_m}.
	\label{eq:dt_parabolic}
\end{equation}

For $m=1$ the equation is discretised in a similar way but for the addition of the source term $\mathcal{F}$. We refer the reader to the next subsection for discretisation methods of the integral contained in it.

Finally, for $m=N$ the equation is actually an ODE, which we approximate by the explicit Euler formula:
$$ (u_N)^{n+1}_{i,j}=(u_N)^n_{i,j}+\frac{\Delta{t}}{2\epsilon}\sum_{\substack{h+k\geq N \\ h,\,k<N}}a_{h,k}(u_h)^n_{i,j}(u_k)^n_{i,j}. $$

\subsection{Discretisation of the equation for $f$}
In the interval $[0,\,1]$, which constitutes the domain of the variable $a$, we introduce a Finite Volume partition made of $N_a$ cells of the form $[a_{k-1/2},\,a_{k+1/2}]$ with central point $a_k=\left(k-\frac{1}{2}\right)\Delta{a}$, where $\Delta{a}=\frac{1}{N_a}$. The cell index $k$ runs from $1$ to $N_a$. Then we discretise the first equation in~\eqref{complete system} using again a fractional step procedure in time.

First, we solve the homogeneous transport part by means of the push-forward scheme introduced by e.g.,~\cite{piccoli2011ARMA,tosin2011NHM}, which is particularly suited to deal with non-local fluxes. Denoting $f^n_{i,j,k}\approx f(x_{i,j},\,a_k,\,t_n)$, we have:
\begin{equation}
	f^\ast_{i,j,k}=f^n_{i,j,k}-\frac{\Delta{t}}{\Delta{a}}\left(
		f^n_{i,j,k}\vert v^n_{i,j,k}\vert-f^n_{i,j,k-1}(v^n_{i,j,k-1})^+
		-f^n_{i,j,k+1}(v^n_{i,j,k+1})^-\right),
	\label{eq:pushfwd}
\end{equation}
where $v^n_{i,j,k}\approx v(x_{i,j},\,a_k,\,t_n)$ indicates an approximation of the deterioration rate of the neurons and $(\cdot)^-$ is the negative part ($x^-:=\max\{0,\,-x\}$). Here we compute $v^n_{i,j,k}$ by approximating the integral contained in the expression~\eqref{velocity bis} via a zeroth order Euler formula and then adding the expression~\eqref{mathcal S}:
$$ v^n_{i,j,k}=\sum_{h=1}^{N_a}\mathcal{G}(x_{i,j},\,a_k,\,a_h)f^n_{i,j,h}\Delta{a}
	+C_\mathcal{S}(1-a_k)\left(\sum_{m=1}^{N-1}m(u_m)^n_{i,j}-\overline{U}\right)^+. $$
If the form $\mathcal{G}(x,\,a,\,b)=C_\mathcal{G}(b-a)^+$ is used then in the formula above we simply have $\mathcal{G}(x_{i,j},\,a_k,\,a_h)=C_\mathcal{G}(a_h-a_k)^+$.

The stability of the scheme~\eqref{eq:pushfwd} requires that the grid steps $\Delta{a},\,\Delta{t}$ be linked by the following CFL condition:
\begin{equation}
	\Delta{t}\leq\frac{\Delta{a}}{\max\limits_{i,\,j,\,k}\vert v^n_{i,j,k}\vert}.
	\label{eq:dt_CFL}
\end{equation}

Second, we update the values $f^\ast_{i,j,k}$ by including the jump process:
$$ f^{n+1}_{i,j,k}=f^\ast_{i,j,k}+\eta\Delta{t}\left(\sum_{h=1}^{N_a}P^k_hf^\ast_{i,j,h}\Delta{a}-f^\ast_{i,j,k}\right)\chi^n_{i,j}, $$
where we have denoted $P^k_h:=P(a_h\to a_k)$ and $\chi^n_{i,\,j}:=\chi(x_{i,j},\,t_n)$.

\subsection{Final choice of the time step}
On the whole, the time step of the complete numerical scheme has to comply with both the parabolic and the hyperbolic constraints~\eqref{eq:dt_parabolic},~\eqref{eq:dt_CFL}, respectively. Therefore, it is ultimately chosen as:
$$ \Delta{t}\leq\min\left\{\frac{\epsilon}{4}\cdot\frac{\min\{\Delta{x}^2,\,\Delta{y}^2\}}{\max\limits_{1\leq m\leq N-1}d_m},\,\frac{\Delta{a}}{\max\limits_{i,\,j,\,k}\vert v^n_{i,j,k}\vert}\right\} $$
at each time iteration of the numerical scheme.

\subsection{Computing physiological indicators}
Several macroscopic (aggregate) quantities can be computed out of the results of model~\eqref{complete system}. In the next Section~\ref{sec:num_res} the outputs of the simulations will be discussed in terms of a few of such quantities, which can be directly compared with real clinical images and known qualitative time evolution of Alzheimer's disease.

The \emph{macroscopic distribution of neuron malfunctioning} $A=A(x,\,t)$ is computed over the cerebral domain $\Omega$ as the local average of the degree of malfunctioning $a$:
$$ A(x,\,t):=\int_0^1 af(x,\,a,\,t)\,da, $$
which is numerically approximated as
$$ A(x_{i,j},\,t_n)\approx A^n_{i,j}=\sum_{k=1}^{N_a}a_kf^n_{i,j,k}. $$

Following~\cite{jack_et_al_2}, we relate then the \emph{local brain atrophy} $\phi(x,\,t)$ to the average neuron malfunctioning $A$ as:
$$ \phi(x,\,t):=\max\left\{0,\,\frac{A(x,\,t)-A_0}{1-A_0}\right\}, $$
$A_0\in (0,\,1)$ being a threshold of malfunctioning over which the brain is considered locally atrophic. The corresponding numerical approximation is
$$ \phi^n_{i,j}=\max\left\{0,\,\frac{A^n_{i,j}-A_0}{1-A_0}\right\}. $$
Next we define the \emph{global brain atrophy} in time $\Phi=\Phi(t)$ as the average of $\phi$ over the whole domain $\Omega$, i.e.,
$$ \Phi(t):=\frac{1}{\vert\Omega\vert}\int_\Omega\phi(x,\,t)\,dx, $$
$\vert\Omega\vert$ denoting the area of $\Omega$, which is numerically approximated as
$$ \Phi^n:=\frac{1}{\vert\Omega\vert}\sum_{i=0}^{N_x-1}\sum_{j=0}^{N_y-1}\phi^n_{i,j}\Delta{x}\Delta{y}, $$
In this formula we conventionally consider $\phi^n_{i,j}=0$ if the grid point $x_{i,j}$ does not belong to the domain $\Omega$, i.e., if it is a point inside the cerebral ventricles.

The \emph{total concentration of soluble amyloid} $U_S=U_S(t)$ in the brain occipital region, to be related to the A$\beta$ concentration found in the cerebrospinal fluid by clinical exams (CSF A$\beta$), is given by:
$$ U_S(t):=\frac{1}{\vert\hat{\Omega}\vert}\int_{\hat{\Omega}}\sum_{m=1}^{N-1}mu_m(x,\,t)\,dx, $$
where $\hat{\Omega}\subset\Omega$ is a subdomain located in the bottom part of $\Omega$, entirely contained in the region below the cerebral ventricles. Assuming for simplicity that it is a rectangle as well, whose grid coordinates are comprised between two indexes $0<\hat{i}_1<\hat{i}_2<N_x$ in the $x$-direction and between $j=0$ and $j=\hat{j}>0$ in the $y$-direction, we obtain the numerical values of $U_S$ as:
$$ U^n_S:=\frac{1}{\vert\hat{\Omega}\vert}\sum_{i=\hat{i}_1}^{\hat{i}_2-1}\sum_{j=0}^{\hat{j}-1}
	\sum_{m=1}^{N-1}m(u_m)^n_{i,j}\Delta{x}\Delta{y}=\frac{1}{(\hat{i}_2-\hat{i}_1)\hat{j}}
		\sum_{i=\hat{i}_1}^{\hat{i}_2-1}\sum_{j=0}^{\hat{j}-1}\sum_{m=1}^{N-1}m(u_m)^n_{i,j}, $$
where we have used that $\vert\hat{\Omega}\vert=(\hat{i}_2-\hat{i}_1)\hat{j}\Delta{x}\Delta{y}$.

Finally, the \emph{average quantity of brain A$\beta$ deposits} in time is:
$$ U_N(t):=\frac{1}{\vert\Omega\vert}\int_\Omega Nu_N(x,\,t)\,dx, $$
which is naturally discretised as
$$ U_N(t_n)\approx U^n_N=\frac{1}{\vert\Omega\vert}\sum_{i=0}^{N_x-1}\sum_{j=0}^{N_y-1}
	N(u_N)^n_{i,j}\Delta{x}\Delta{y} $$
by letting conventionally $(u_N)^n_{i,j}=0$ if $x_{i,j}\not\in\Omega$ (inside the cerebral ventricles).

\section{Numerical results}
\label{sec:num_res}
\begin{figure}[!ht]
\centering
\includegraphics[width=9cm]{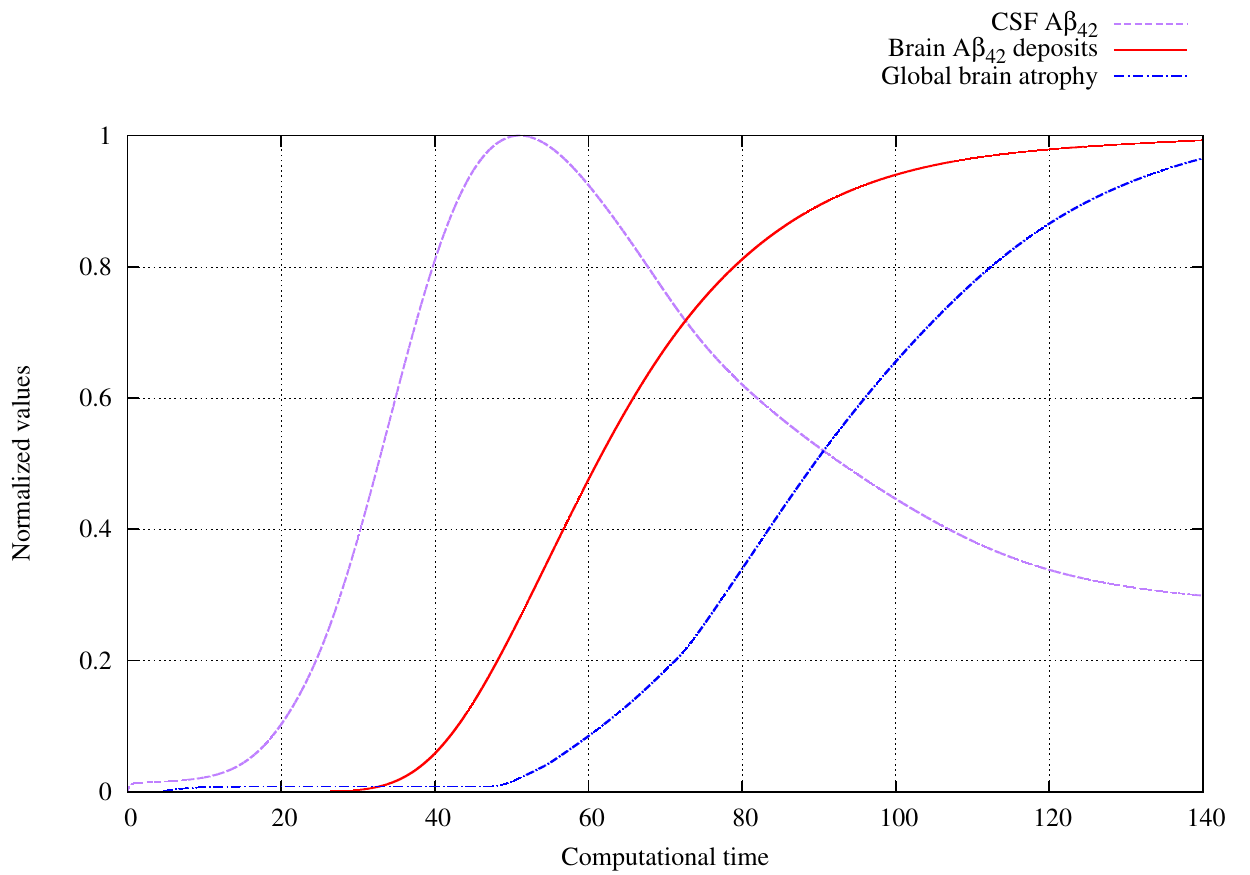}
\caption{Graph for the following constants: $\beta=1$, $D=0.01$, $\alpha=10$, $\eps=0.1$, $T=100$, $N=50$, $C_{\mathcal G}=0.1$, $C_{\mathcal S}=0.001$, $C_{\mathcal F}=10$, $r_0=0.0$, $\overline{U}=0.1$, $\mu_0=0.01$, $\eta=1$, and $\sigma_m=1/m$.}
\label{fig:beta=1}
\end{figure}

To begin with, we provide a typical output of the numerical simulations. In Figure \ref{fig:beta=1} we plot the evolution of three crucial biomarkers of AD (as a function of the computational  time): 
\begin{itemize}
\item[$-$] the CSF A$\beta_{42}$ (purple dashed curve);
\item[$-$] the average quantity of brain A$\beta_{42}$ deposits (red solid curve);
\item[$-$]  the global brain atrophy (blue dash-dot curve).
\end{itemize}
All curves are normalised to their maxima. The values of the constants used in the simulation are specified in the figure caption.

The level of A$\beta_{42}$-deposition (red solid curve) grows rapidly, reaches its maximum, and then stabilises. The purple dashed curve, corresponding to CSF-A$\beta_{42}$, decreases after having reached a peak. The blue dash-dot curve corresponds to the brain atrophy  and increases in time as expected. The graphs in  Figure~\ref{fig:beta=1} can be well illustrated by the following quote from~\cite{jack_et_al_1}:
\begin{quotation}
The initiating event in AD is related to abnormal processing of $\beta$-amyloid peptide, ultimately leading to formation of A$\beta$ plaques in the brain. This process occurs while individuals are still cognitively normal. Biomarkers of brain $\beta$-amyloidosis are reductions in CSF A$\beta_{42}$ and increased amyloid PET tracer retention. After a lag period, which varies from patient to patient, neuronal dysfunction and neurodegeneration become the dominant pathological processes. Biomarkers of neuronal injury and neurodegeneration are increased CSF tau and structural MRI measures of cerebral atrophy. Neurodegeneration is accompanied by synaptic dysfunction, which is indicated by decreased fluorodeoxyglucose uptake on PET. We propose a model that relates disease stage to AD biomarkers in which A$\beta$ biomarkers become abnormal first, before neurodegenerative biomarkers and cognitive symptoms, and neurodegenerative biomarkers become abnormal later, and correlate with clinical symptom severity.
\end{quotation}

\begin{figure}[!ht]
\centering
\includegraphics[width=9cm]{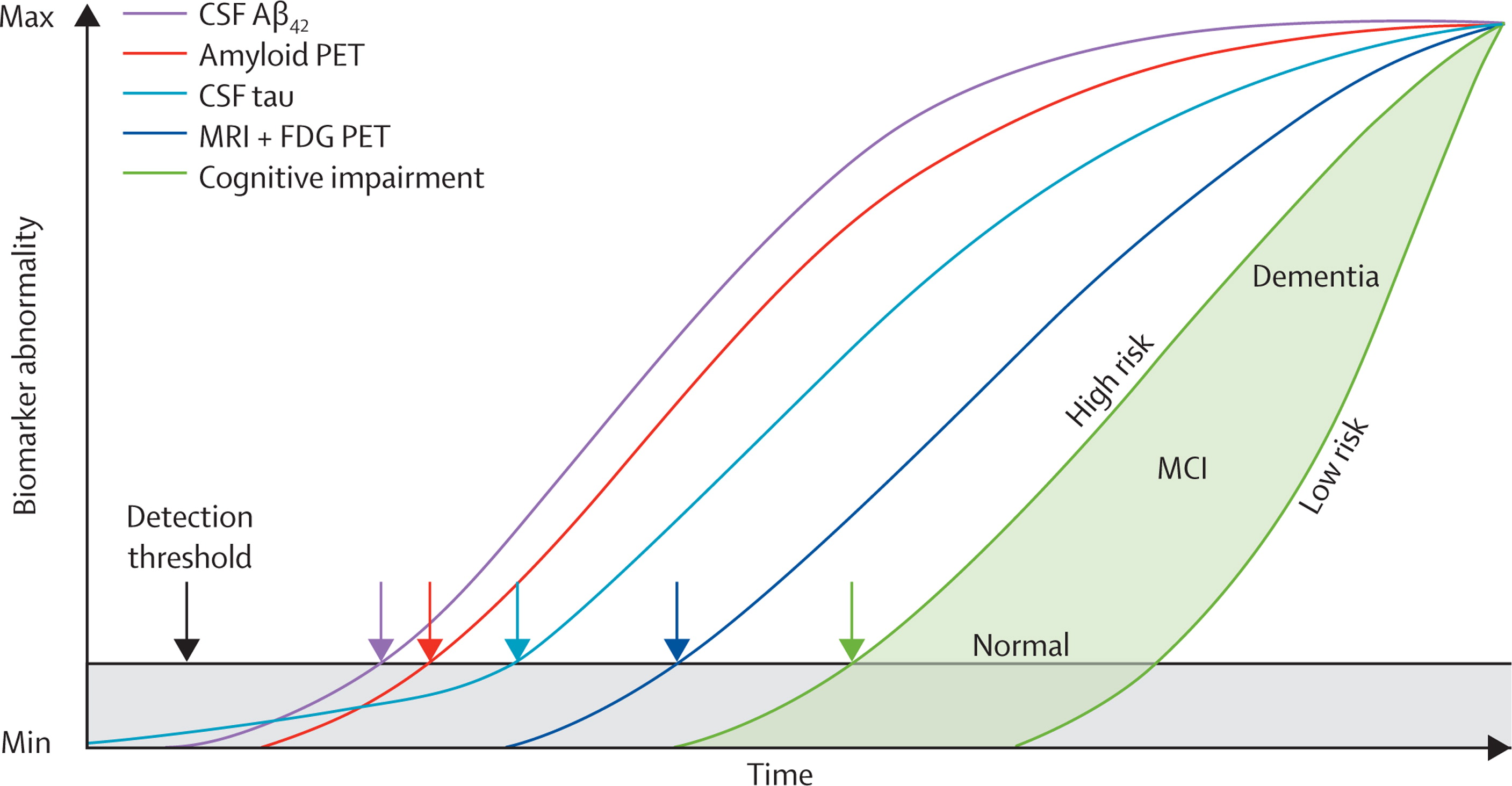}
\caption{Fig. 6 reproduced from~\cite{jack_et_al_2} with kind permission of the publisher.}
\label{fig:jack2013}
\end{figure}

\begin{figure}[!ht]
\centering
\includegraphics[width=9cm]{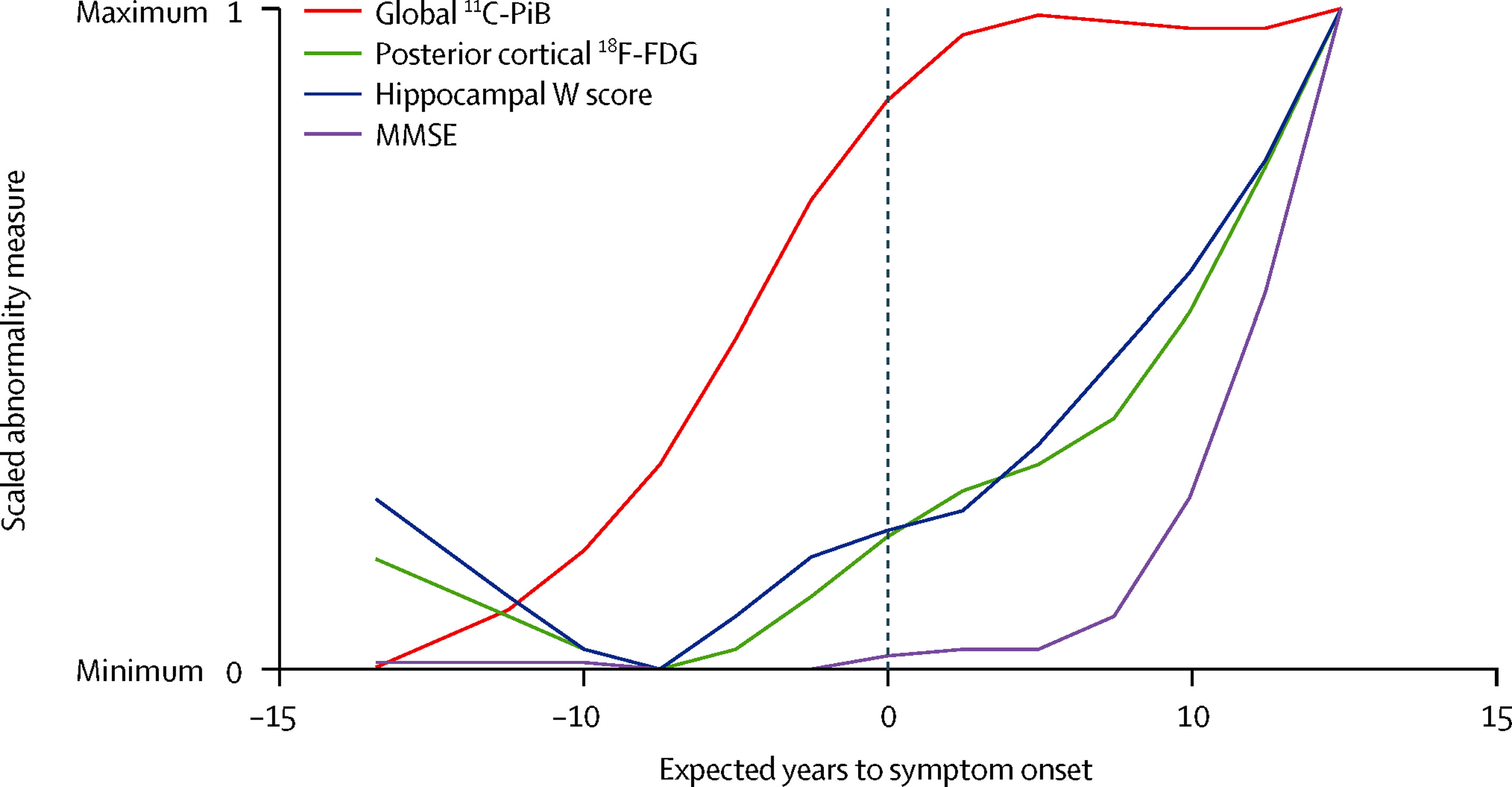}
\caption{Fig. 4 reproduced from~\cite{yau2015} with kind permission of the publisher.}
\label{fig:yau}
\end{figure}

The plots we obtain should be compared with the clinical graphs in ~\cite{jack_et_al_2}, ~\cite{yau2015}, ~\cite{bateman}, and with the data of ~\cite{reiman2} and ~\cite{reiman1}. For the reader's convenience we reproduce here a picture from ~\cite{jack_et_al_2}, see Figure \ref{fig:jack2013}, and a picture from ~\cite{yau2015}, see Figure \ref{fig:yau}.

There is a satisfactory agreement between the plots of the qualitative temporal behaviour of the biomarkers and those obtained from clinical data. Observe that not only the shapes of the curves are comparable (CSF A$\beta$ corresponds to CSF A$\beta_{42}$, Brain A$\beta$ deposits correspond to Amyloid PET and Global brain atrophy corresponds to MRI + FDG PET)), but also the temporal order of the events is in good agreement with clinical data.

Obviously the details of the numerical output depend on the choice of the constants used in the mathematical model. Performing a considerable amount of numerical runs with different values of the constants in the model, we have reached the conclusion that, at least qualitatively, the behaviour of the solutions does not depend on the precise choice of those constants, as long as their variation is restricted to reasonable ranges. In other words, the values of the constants taken in Figure \ref{fig:beta=1} can be considered as an indication for the order of magnitude. For example, the longitudinal graphs of the biomarkers CFS-A$\beta$, brain A$\beta$ deposits and brain atrophy are - in this sense - qualitatively stable under variations of $C_{\mathcal S}$, $C_{\mathcal F}$, $C_{\mathcal G}$ and $\alpha$.

It is particularly instructive to consider the constants $\overline U$ in \eqref{mathcal S} and $\beta$ in \eqref{eq:bc_Robin}. We recall that $\overline U$ is a threshold value for the minimal amount of toxic A$\beta$ necessary to damage neurons (see~\eqref{mathcal S}). In Figure \ref{fig:beta=1} we have used the value $\overline{U}=0.1$, but if we make it considerably larger, for example $\overline U=1$ (the remaining constants are unchanged), then the threshold  becomes so high that the illness does not develop at all.

\begin{figure}[!ht]
\centering
\includegraphics[width=9cm]{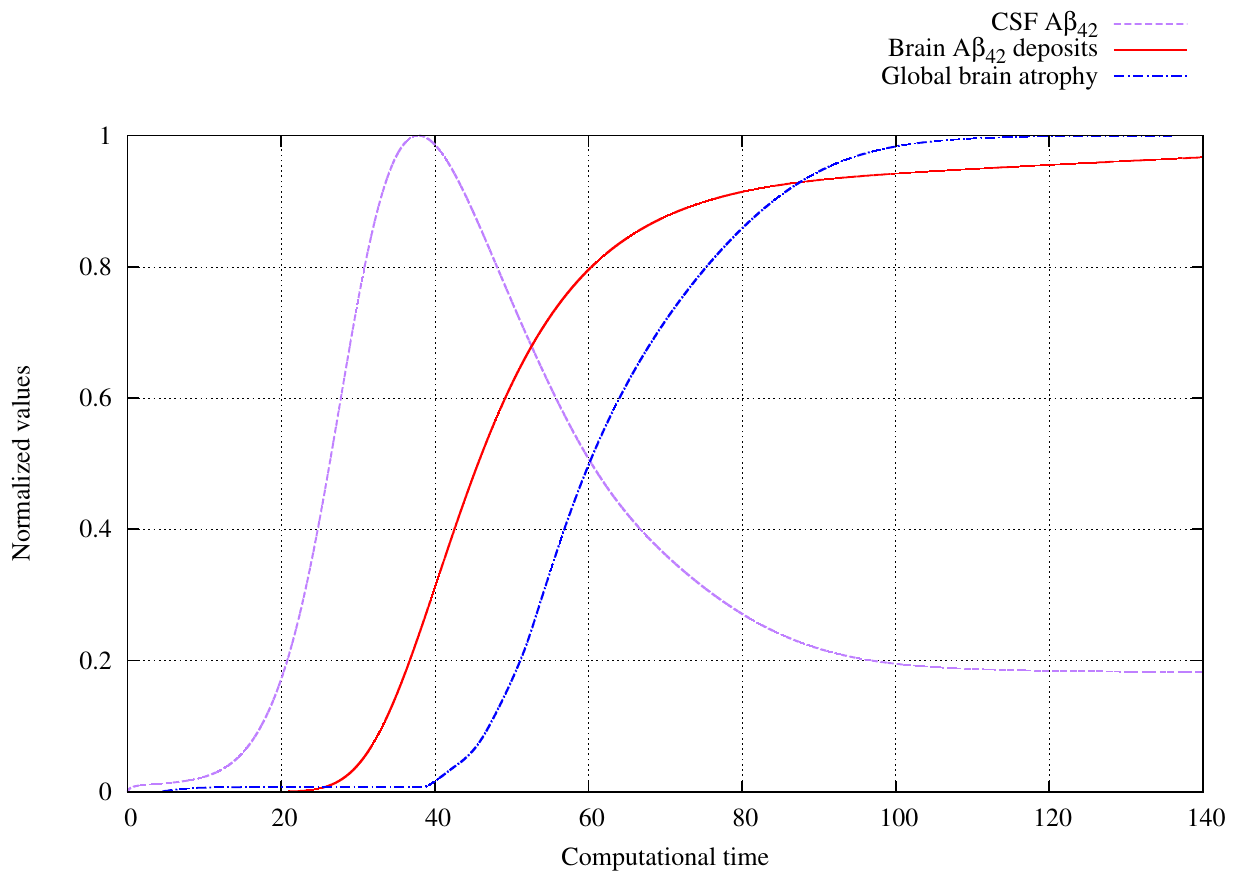}
\caption{Graph for the following constants: $\beta=0.01$, $D=0.01$, $\alpha=100$, $\eps=0.1$, $T=100$, $N=50$, $C_{\mathcal G}=0.1$, $C_{\mathcal S}=0.001$, $C_{\mathcal F}=10$, $r_0=0.0$, $\overline{U}=0.1$,$\mu_0=0.01$, $\eta=1$, and $\sigma_m=1/m$.}
\label{fig:beta=0.01}
\end{figure}

The constant $\beta$ enters the model through condition~\eqref{eq:bc_Robin} at the boundary of the cerebral ventricles. Smaller values of $\beta$ mean that less A$\beta$ is removed from the CSF through the choroid plexus. Figure \ref{fig:beta=0.01} shows what happens if we change it into $\beta=0.01$: the three curves are moved to the left and become steeper: the illness starts earlier and develops faster. Recalling that in Figures \ref{fig:beta=1} and \ref{fig:beta=0.01} we have plotted values which are normalised with respect to their maximal values, one could wonder how the latter ones depend on $\beta$. It turns out that the maximal values of CSF A$\beta$ and the brain atrophy are essentially independent of $\beta$. The A$\beta$ deposits (the plaques) however increase by a factor 6 if $\beta$ is changed from 1 to 0.01. This result is compatible with our modelling Ansatz (in accordance with the medical literature) that plaques are not toxic  (even healthy brains may contain plaques).

The comparison of the cases $\beta=1$ and $\beta=0.01$ becomes even clearer when we create spatial plots of $f$ and of the distribution and density of the cerebral plaques at fixed computational times $t=T$. The plots of $f$ at different times are meant to be compared with FDG-PET images (see e.g.,~\cite{reiman2}). More precisely, we take a schematic image of a transverse section of the brain and attribute different colors (varying from red to blue) to those parts of the brain where probabilistically the level of malfunctioning lies in different ranges. As in the FDG-PET, the red corresponds to a healthy tissue. Here AD originates only from the hippocampus and propagates, at the beginning, along privileged directions (such as those corresponding to denser neural bundles) mimicked by two triangles. 

\begin{figure}[!ht]
\centering
\includegraphics[width=0.45\textwidth]{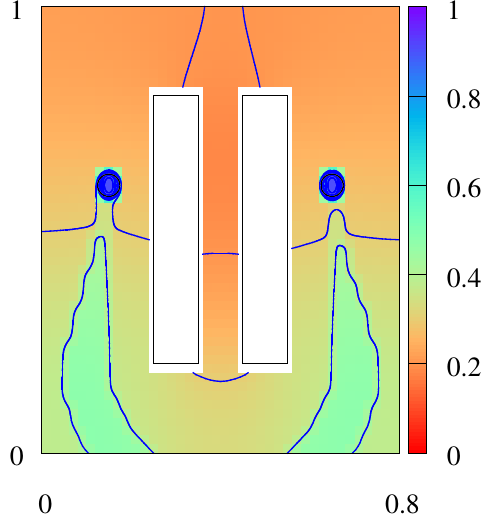}
\qquad
\includegraphics[width=0.45\textwidth]{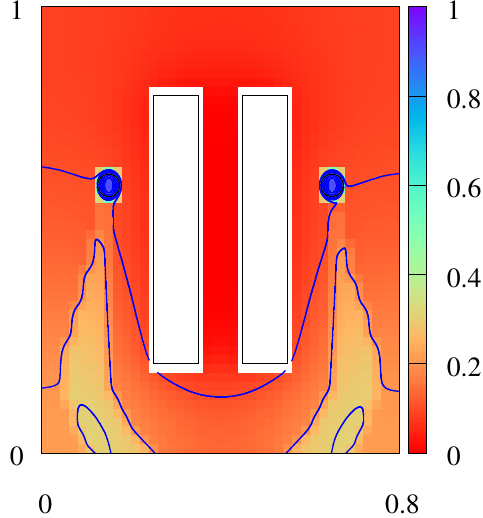}
\caption{Neuron malfunctioning: $\beta=0.01$ (left), $\beta=1$ (right), $T=34$.}
\label{fig:diseaseT=34}
\end{figure}

\begin{figure}[!ht]
\centering
\includegraphics[width=0.45\textwidth]{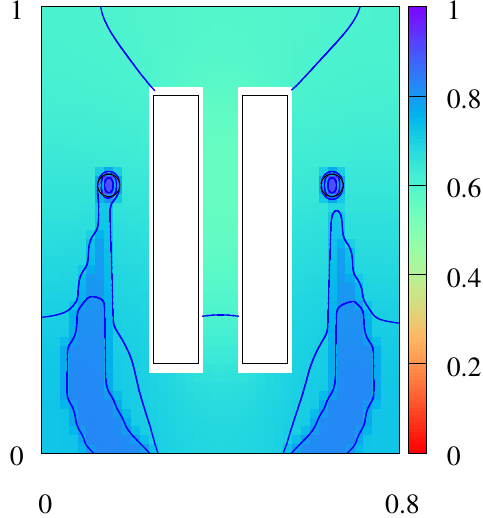}
\qquad
\includegraphics[width=0.45\textwidth]{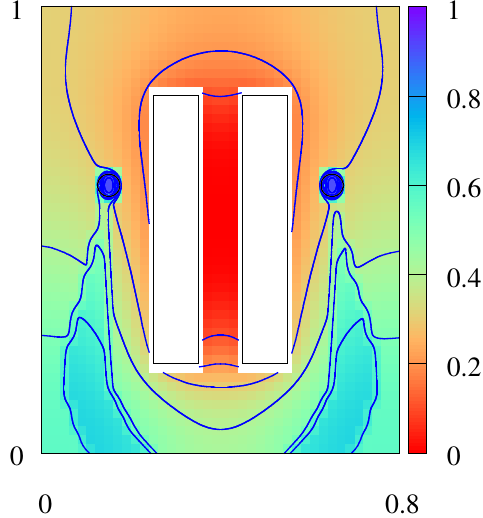}
\caption{Neuron malfunctioning: $\beta=0.01$ (left), $\beta=1$ (right), $T=52$.}
\label{fig:diseaseT=52}
\end{figure}

In Figures \ref{fig:diseaseT=34} and \ref{fig:diseaseT=52} we compare plots of $f$ at, respectively, times $T=34$ and $T=52$ for the two different values of $\beta =0.01$ and $\beta=1$. Figures \ref{fig:diseaseT=34} and \ref{fig:diseaseT=52} do not only confirm the temporal acceleration of the development of the illness for smaller values of $\beta$, but also show that the spatial pattern and heterogeneity become less evident as $\beta$ becomes smaller. Since experimental data suggest a strong spatial heterogeneity of the illness, this could indicate the potential importance of the removal of $A\beta$ through the choroid plexus to slow down the temporal development of AD.

\begin{figure}[!ht]
\centering
\includegraphics[width=0.45\textwidth]{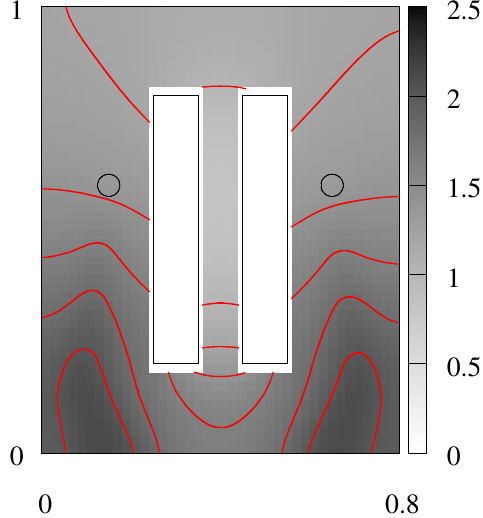}
\qquad
\includegraphics[width=0.45\textwidth]{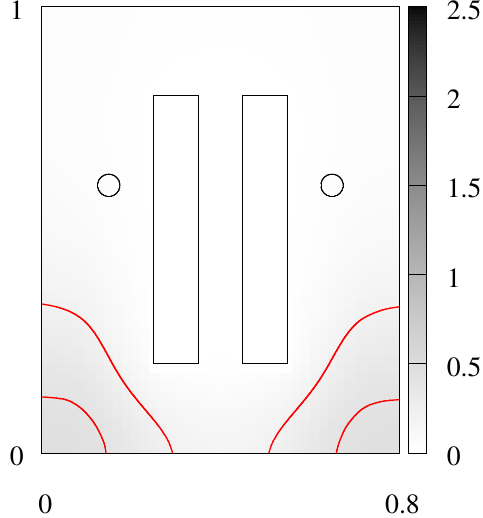}
\caption{Density of plaques for $\beta=0.01$ (left), $\beta=1$ (right), $T=52$.}
\label{fig:placcheT=52}
\end{figure}

In Figure \ref{fig:placcheT=52} we plot the plaques' distribution for the two different values of $\beta =0.01$ and $\beta=1$ and at $T=52$. This figure confirms the strong increase of the plaques when $\beta$ becomes smaller.

We stress that, though our images represent \emph{a mean value} of brain activity instead of a single patient's brain activity, still they show a good agreement with clinical neuroimaging: compare Figures \ref{fig:diseaseT=34} and \ref{fig:diseaseT=52} with Figure \ref{pet} below.

\begin{figure}[!ht]
\centering
\includegraphics[width=6.5cm]{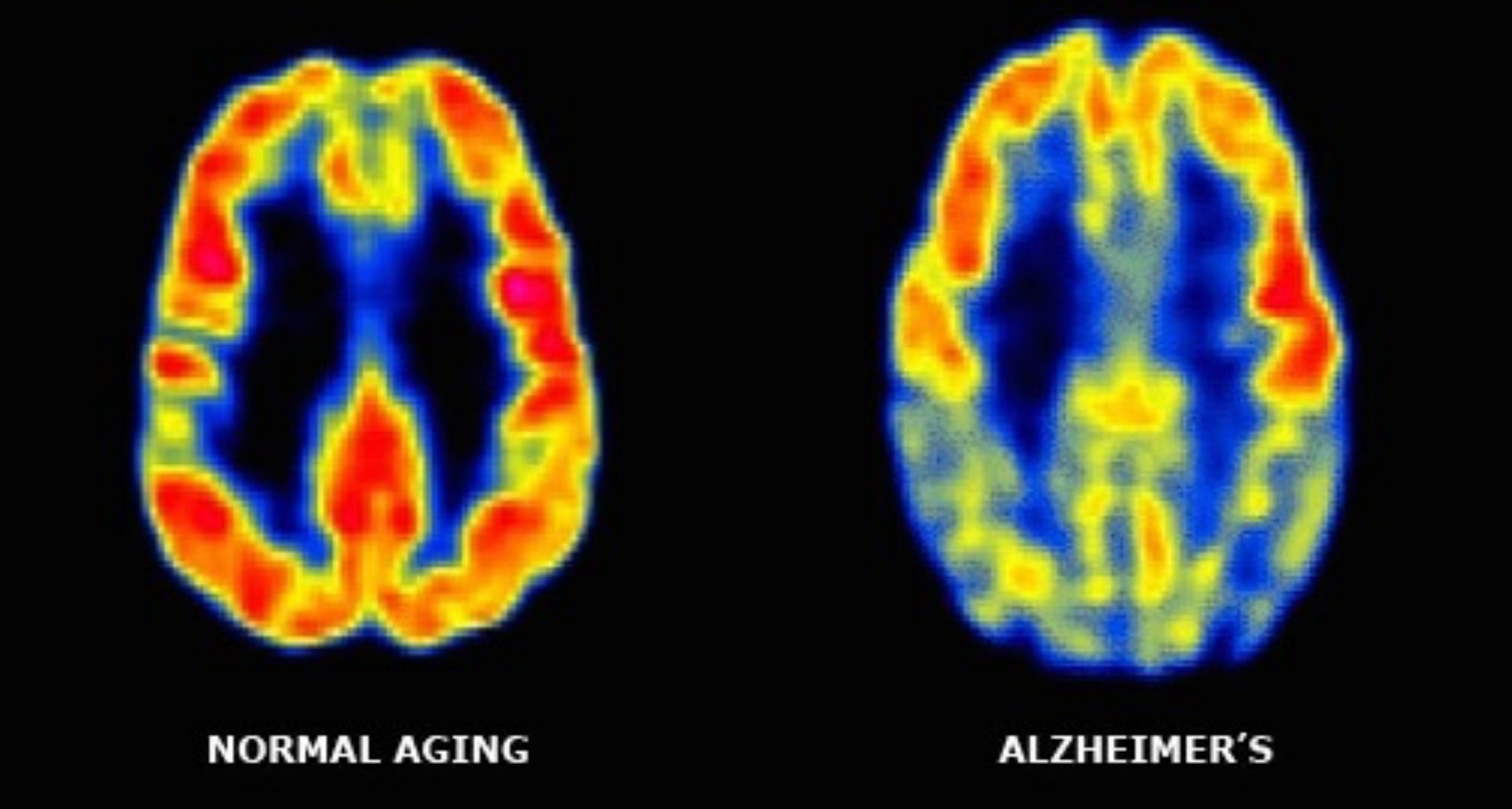}
\caption{FDG -PET images showing patterns of metabolic activity: an elderly individual with no dementia (left) and with AD (right).
Reproduced from~\cite{miller2006RR} with permission.}
\label{pet}
\end{figure}

\begin{figure}[!ht]
\centering
\includegraphics[width=6.5cm]{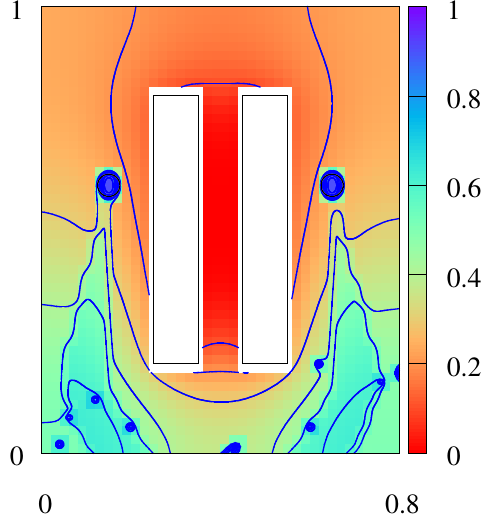}
\caption{Neuron disease with random sources with $\beta=1$ at $T=47$}
\label{fig:disease_sorgenti_casuali}
\end{figure}

Looking for more realistic images, we have to take into account randomness of the spatial distribution of the sources of the disease. For example, we have performed some runs where the AD does not only originate from the hippocampus, but also from several sources of A$\beta$ randomly distributed in the occipital part of the brain. We report the outputs of such runs in Figure~\ref{fig:disease_sorgenti_casuali}. The random distributed sources appear as  the small blue spots.

\section{Discussion and future research directions}
We have presented a new mathematical model for the onset and evolution of AD. The model is characterised by a high level of flexibility, which potentially allows one to simulate different modelling hypotheses and compare them with clinical data. In fact, the model provides a flexible tool to test in the future alternative hypotheses on the evolution of the disease. In the paper we have chosen  some specific aspects of the illness, such as the aggregation, diffusion and removal of $A\beta$, the possible spread of neuronal damage in the neural pathway, and, to describe the onset of AD, a random neural deterioration mechanism. Numerical simulations are compared with clinical data and, although oversimplified and restricted to a 2-dimensional rectangular section of the brain, they are in good qualitative agreement with the spread of the illness in the brain at various stages of its evolution. In particular, our model  captures the cerebral damage in the early stage of MCI. 

There are multiple future research developments in quite different directions, each of which requires substantial research efforts. We mention some of them.

Further development of the model is needed and should be carefully guided by clinical data. The constants appearing in the equations should be well calibrated to optimise quantitative agreement with clinical data. Simulations should become more realistic, in a three-dimensional domain which matches the geometric characteristics of the brain.

The true challenge in AD research is a breakthrough which allows one to develop effective therapies to stop or slow down the evolution of AD, possibly in an early stage of the illness. Also effective mathematical models can give a contribution in this direction. For example, a certain sensibility of the numerical output to the value of the constant $\beta$ in \eqref{eq:bc_Robin}, which models the removal of $A\beta$ through the choroid plexus, spontaneously leads to the question whether dialysis-mechanisms can be introduced to enhance $A\beta$-removal artificially. Most probably, a serious answer to this question requires, in addition to a detailed comparison with clinical data, a more refined modelling of the removal which takes into account the transport of soluble A$\beta$ by the cerebral fluid. 
    
Finally, some mathematical effort is necessary to check the mathematical correctness (well-posedness) of the model.

\section*{Acknowledgments}
B.~F. and M.~C.~T. are supported by a grant of the University of Bologna (Ricerca Fondamentale Orientata). B.~F. and M.~C.~T. are supported by GNAMPA (Gruppo Nazionale per l'Analisi Matematica, la Probabilit\`{a} e le loro Applicazioni) of INdAM (Istituto Nazionale di Alta Matematica), Italy. A.~T. acknowledges that this work has been written within the activities of GNFM (Gruppo Nazionale per la Fisica Matematica) of INdAM (Istituto Nazionale di Alta Matematica), Italy.

\bibliographystyle{amsplain}
\bibliography{BmFbMnTmcTa-alzheimer}

\end{document}